\documentclass[superscriptaddress,prl,twocolumn,amssymb,aps]{revtex4}

\usepackage{amsmath,bm}
\usepackage{graphicx}
\usepackage[latin1]{inputenc}
\usepackage[T1]{fontenc}
\usepackage{latexsym}
\usepackage{times}
\usepackage{float}
\usepackage{soul}
\usepackage{url}
\usepackage{amsmath,amsfonts,amsthm,bm} % Math packages

\begin{document}

%\title{Modelling the detection and characterisation of correlated light with a photon counting camera}
%\title{Modelling the detection of biphotons using an EMCCD camera and their ratio to spatially uncorrelated photons}
%\title{Modelling and characterisation of biphotons as detected by a photon-counting camera and quantitative verification by experiment}
%\title{The modelling and characterisation of biphotons detected by a photon-counting camera and a quantitative verification by experiment}
%\title{A model for quantum imaging and sensing experiments based on the detection of correlated biphotons using a photon-counting camera}
\title{A model for the detection of spatially correlated biphotons using a photon-counting camera}
\author{Ermes Toninelli}
\author{Paul-Antoine Moreau}
\author{Thomas Gregory}
\author{Miles J. Padgett}

\email[Corresponding author: ]{miles.padgett@glasgow.ac.uk}
\affiliation{School of Physics and Astronomy, University of Glasgow, G12 8QQ, UK}

\date{\today}

\begin{abstract}
\noindent 
%In the paper presented here we describe a model that allows the generation of images simulated as though acquired by a photon counting camera. Such a model is helpful to simulate and characterise the detection of quantum correlations by such a camera. We derive the expected values of quantities characteristic of the quality of the correlations detected in a given experiment. We also show good agreement between the model predictions and experimental measurements.
Spontaneous downconversion is a versatile source for correlated biphotons that has been employed in many quantum sensing and imaging experiments. Spatially-resolved photon-counting detectors allow to access a large number of modes, posing the challenge of an accurate description of such systems. We propose a simple model to generate images as though acquired by a photon-counting camera, and allow to simulate and characterise the detection of quantum correlations. We derive quantitative parameters characteristic of the spatial correlations for a given experiment, comparing the images produced by our model to the frames acquired by an electron-multiplying CCD camera. Moreover we accurately predict the decreased detection of spatially correlated biphotons caused by introducing a variable amount of optical loss after the nonlinear crystal, even though the total number of detected events is kept constant, showing excellent agreement between model and experiment.
\end{abstract}

\maketitle

The detection of quantum correlations using single-photon sensitive cameras has attracted a lot of attention in recent years. Previous efforts employed avalanche photodiodes to scan for correlated photon-events or few-detectors' arrays~\cite{pittman_OpticalImagingMeans_1995,leach_QuantumCorrelationsPosition_2012}. 
The high dimensionality accessible to quantum states of light through the spatial domain makes efficient characterisation of high dimensional entanglement possible using such devices~\cite{edgar_ImagingHighdimensionalSpatial_2012, Devaux2012,moreau_EinsteinPodolskyRosenParadoxTwin_2014, reichert_MassivelyParallelCoincidence_2018,reichert_QualitySpatialEntanglement_2017}. Such characterisation of correlations can find applications in quantum information, sensing and imaging schemes, that can benefit from such high dimensionality, thus enabling a greater quantum advantage by harnessing the increasing complexity of such quantum states. 

The range of applications includes fundamental tests of quantum mechanics~\cite{howell_RealizationEinsteinPodolskyRosenParadox_2004}, quantum information processing and communications~\cite{tasca_ContinuousvariableQuantumComputation_2011}, quantum ghost imaging~\cite{morris_ImagingSmallNumber_2015,moreau_ResolutionLimitsQuantum_2018,moreau_ExperimentalLimitsGhost_2018}, sub-shot-noise sensing~\cite{jedrkiewicz_DetectionSubShotNoiseSpatial_2004,blanchet_MeasurementSubShotNoiseCorrelations_2008,toninelli_SubshotnoiseShadowSensing_2017}, as well as noise~\cite{brida_ExperimentalRealizationSubshotnoise_2010}, contrast~\cite{izdebski_QuantumCorrelationsMeasured_2013}, and resolution enhanced imaging~\cite{schwartz_ImprovedResolutionFluorescence_2012,toninelli_ResolutionenhancedQuantumImaging_2019,tenne_SuperresolutionEnhancementQuantum_2019} using quantum correlations. Therefore understanding systems that utilise entangled photon pairs and which require or would be improved by the detection of the transverse positions of said photons using two-dimensional detector arrays is of great interest. In terms of modelling, the detecting behaviour of both intensified and electron-multiplying cameras have been investigated focussing on the noise performance, the optimisation of the signal-to-noise ratio, photon-counting strategies, and a Bayesian description of these detectors~\cite{robbins_NoisePerformanceElectron_2003,tasca_OptimizingUseDetector_2013,lantz2014optimizing,doi:10.1046/j.1365-8711.2003.07020.x,lantz_MultiimagingBayesianEstimation_2008}. Additionally, the related absolute calibration in terms of detection efficiency for EMCCD and ICCD cameras can be found in the following works~\cite{avella_AbsoluteCalibrationEMCCD_2016,qi_AutonomousAbsoluteCalibration_2016a}.\\

Here we report a general model that can accurately reproduce the spatially resolved detection of degenerate twin-photons produced by spontaneous parametric downconversion. The model produces binary frames (i.e. photon-counted), as those detected by a single-photon sensitive camera placed in either the near- or far-field of a type-I nonlinear crystal. The generality of this model means it can be easily adapted to also describe the detection of twin-photons produced by a type-II crystal, as well as various phase-matching conditions. Additionally, the model can be used to extract quantitative information about the strength of the correlations, allowing to extract the number of spatially correlated and uncorrelated photons from acquired binary frames. The effectiveness of the model is demonstrated by comparing its generated frames to those acquired using an EMCCD cameras. We show that under constant illumination, it is possible to successfully predicting the decreased detection of spatially correlated biphotons as a consequence of increased optical loss. This accurate and simple model can therefore be used to simulate a variety of quantum imaging and sensing experiments, on a photon by photon basis.

\section{Detecting the signature of quantum correlations: the correlation function}
In this section we show that the intensity correlation function for a spatially resolved detector and the cross-correlation function of a pixelated image with a 180-degree rotated copy of itself are mathematically equivalent. Specifically, we show that in the presence of spatially correlated events (i.e. biphotons) both functions are made of two contributions: a term representing the presence of quantum correlations; and a term representing accidental coincidences. We then express the isolated contribution due to detected photon-pairs, allowing to estimate the number of photons that are actually detected with their twins.

In the present context we are concerned with the detection of spatial anticorrelations between downconverted photons in the far-filed of a nonlinear crystal. The quantum theory of photo-detection~\cite{RevModPhys.71.1539} gives the following expression for the expected intensity correlation function $C(\bm{\rho_1},\bm{\rho_2})$ obtained between two pixels of coordinates $\bm{\rho_1}$ and $\bm{\rho_2}$, according to Eq.~\ref{eq:quantumDetection}:
\begin{eqnarray}\label{eq:quantumDetection}
C(\bm{\rho_1},\bm{\rho_2})&=&G^{(2)}(\bm{\rho_1},\bm{\rho_2})+\left\langle\hat{N}_1(\bm{\rho_1})\right\rangle\left\langle\hat{N}_2(\bm{\rho_2})\right\rangle\nonumber\\
&=&\left\langle :\hat{N}_1(\bm{\rho_1})\hat{N}_2(\bm{\rho_2}):\right\rangle+\left\langle\hat{N}_1(\bm{\rho_1})\right\rangle\left\langle\hat{N}_2(\bm{\rho_2})\right\rangle\nonumber\\
&=&\left\langle a^\dag_1a^\dag_2a_1a_2\right\rangle+\left\langle a^\dag_1a_1\right\rangle\left\langle a^\dag_2a_2\right\rangle,
\end{eqnarray}
where the first term $G^{(2)}(\bm{\rho_1},\bm{\rho_2})$ accounts for quantum correlations and the second term $\left\langle\hat{N}_1(\bm{\rho_1})\right\rangle\left\langle\hat{N}_2(\bm{\rho_2})\right\rangle$ corresponds to accidental coincidences due to the mean number of photons detected on each pixel. Here we use $G^{(2)}$, $\hat{N}_1$, and $\hat{N}_2$ to indicate respectively the second order correlation operation and the photon-number operators within the two beams 1 and 2. The notation $:\text{ }:$ indicates normal ordering, according to which all creation operators ($a^\dag_1$ and $a^\dag_2$) are placed to the left of the annihilation operators ($a_1$ and $a_2$).
%We can therefore distinguish two contributions to the correlation function, the first term $G^{(2)}(\bm{\rho_1},\bm{\rho_2})$ accounts for quantum correlations while the second term $\left\langle\hat{N}_1(\bm{\rho_1})\right\rangle\left\langle\hat{N}_2(\bm{\rho_2})\right\rangle$ corresponds to the accidental coincidences due to the number of photons detected on each of the pixels.

It is possible to isolate the contribution due to quantum correlations $\tilde{C}(\bm{\rho_1},\bm{\rho_2})\equiv G^{(2)}(\bm{\rho_1},\bm{\rho_2})$, as follows:
\begin{eqnarray}\label{eq:qc_component}
\tilde{C}(\bm{\rho_1},\bm{\rho_2})&=&\left\langle :\delta\hat{N}_1\delta\hat{N}_2:\right\rangle\\
&=&\left\langle :\hat{N}_1\hat{N}_2:\right\rangle-\left\langle\hat{N}_1\right\rangle\left\langle\hat{N}_2\right\rangle\\
&=&C(\bm{\rho_1},\bm{\rho_2})-\left\langle\hat{N}_1\right\rangle\left\langle\hat{N}_2\right\rangle.
\end{eqnarray}

Having formulated the two contributions to the intensity correlation function, we draw the mathematical analogy with the cross-correlation function, thus allowing to identify its spatially correlated and uncorrelated contributions. The spatial cross-correlation function $\mathcal{C}(\bm{\rho})$ of a frame with a  180-degree rotated copy of itself can be expressed as:
\begin{eqnarray}
\mathcal{C}(\bm{\rho})&=&\sum^{}_{\bm{\rho'}}C(\bm{\rho'},-(\bm{\rho'}+\bm{\rho}))\\
&=&\sum^{}_{\bm{\rho'}}\left\langle \hat{N}(\bm{\rho'})\hat{N}(-(\bm{\rho'}+\bm{\rho}))\right\rangle,
\end{eqnarray}
where the summation applies to each pixel of the original frame. As in the case of the intensity correlation function of Eq.\ref{eq:quantumDetection}, it is possible to isolate two contributions from the spatial cross-correlation function. The first accounts for the presence of quantum correlations and the second accounts for accidental coincidences. We can therefore rewrite this quantity as:
\begin{eqnarray}
\mathcal{C}(\bm{\rho})&=&\sum^{}_{\bm{\rho'}}\left[G^{(2)}\left(\bm{\rho'},-(\bm{\rho'}+\bm{\rho})\right)\right.\nonumber\\& &+\left.\left\langle\hat{N}(\bm{\rho'})\right\rangle\left\langle\hat{N}(-(\bm{\rho'}+\bm{\rho}))\right\rangle\right].
\end{eqnarray}
When the state of the light is composed of photon pairs such as generated by spontaneous parametric down conversion (SPDC light) i.e. in the regime where the number of photons per modes is low, $G^{(2)}\left(\bm{\rho_1},\bm{\rho_2}\right)=N_p \mathcal{P}(\bm{\rho_1},\bm{\rho_2})$, where $\mathcal{P}(\bm{\rho_1},\bm{\rho_2})$ is the joint detection probability of the photon pairs and $N_p$ is the mean number of photon pairs. This result is a signature of a Poissonian statistics that is valid for describing the SPDC process only at low light level when the stimulated process is negligible. On the other hand if the light state is solely composed of single photons, $G^{(2)}\left(\bm{\rho_1},\bm{\rho_2}\right)=0$.\\
By computing the correlation function $\mathcal{C}(\bm{\rho})$ from experimental images we can isolate the contribution $\mathcal{C}_q(\bm{\rho})$ of the photon pairs:
\begin{eqnarray}
\mathcal{C}_q(\bm{\rho})&=&\sum^{}_{\bm{\rho'}}N_p \mathcal{P}(\bm{\rho'},-(\bm{\rho'}+\bm{\rho}))\\&=&\mathcal{C}(\bm{\rho})-\sum^{}_{\bm{\rho'}}\left\langle\hat{N}(\bm{\rho'})\right\rangle\left\langle\hat{N}(-(\bm{\rho'}+\bm{\rho}))\right\rangle
\end{eqnarray}
Experimentally the statistical averaging $\left\langle~~\right\rangle$ can be computed either through a temporal averaging by recording many images or through a purely spatial averaging using a single frame~\cite{lantz2015einstein}. We can observe that the integral of this contribution gives an estimation of the number of photons that are actually detected with their twin:
\begin{eqnarray}
\sum^{}_{\bm{\rho}}\mathcal{C}_q(\bm{\rho})=\sum^{}_{\bm{\rho}}\sum^{}_{\bm{\rho'}}N_p \mathcal{P}(\bm{\rho'},-(\bm{\rho'}+\bm{\rho}))=2 N_p
\end{eqnarray}
The factor of 2 is due to the fact that each pair is taken into account twice when integrating as the same image containing the two photons that the anticorrelated pair consists of. The integral of the correlation peak due to the quantum correlation contribution therefore gives an estimation of the number of detected pairs $N_p$. It will be zero for uncorrelated light.\\

Another interesting feature that can be extracted from the correlation function is the total number of detected events in the frames. We now consider the experimental correlation defined as follows:
\begin{equation}
\mathcal{C}_{exp}(\bm{\rho})=\sum^{}_{\bm{\rho'}} N(\bm{\rho'})N(-(\bm{\rho'}+\bm{\rho}))
\end{equation}
where $N(\bm{\rho})$ is the number of photons detected in a frame on the pixel of coordinate $\bm{\rho}$. $\mathcal{C}_{exp}(\bm{\rho})$ is simply obtained by computing the cross correlation of a frame with itself rotated by $180^{\circ}$.\\
One can now define the integrated correlation :
\begin{equation}
\mathcal{C}=\sum^{}_{\bm{\rho}}C_{exp}(\bm{\rho})=\sum^{}_{\bm{\rho}}\sum^{}_{\bm{\rho'}} N(\bm{\rho'})N(-(\bm{\rho'}+\bm{\rho}))
\end{equation}
Using Fubini's theorem one can show that :
\begin{equation}
\mathcal{C}=\sum^{}_{\bm{\rho'}} N(\bm{\rho'})\sum^{}_{\bm{\rho'}}N(-\bm{\rho'})=N_t^2
\label{eq:fubini}
\end{equation}
The  integral of the correlation function therefore gives the square of the total number of detections in the frame.\\

It can be noted at this point that the mean integrated correlation $\left\langle\mathcal{C}\right\rangle$ will be greater for correlated light than for uncorrelated light for the same mean number of detection events in the frame. It is a result from the fact that $N_t$ will be super-Poissonian in presence of photon pairs:
\begin{eqnarray}
\left\langle\mathcal{C}\right\rangle&=&\left\langle N_t^2\right\rangle=\left\langle(N_s+2 N_p)^2\right\rangle\\&=&\left\langle(N_t)\right\rangle^2+\left\langle(N_s)\right\rangle+4\left\langle(N_p)\right\rangle\\
&=&\left\langle(N_t)\right\rangle^2+\left\langle(N_t)\right\rangle+2\left\langle(N_p)\right\rangle\\
&=&\left\langle(N_t)\right\rangle^2+\left\langle(N_t)\right\rangle+\sum^{}_{\bm{\rho}}\mathcal{C}_q(\bm{\rho})
\end{eqnarray}
Where $N_s$ is the number of singles and we have used the fact that both $N_s$ and $N_p$ follow a Poisson distribution ($\text{var}(N)=N$). The correlation excess corresponds exactly to the contributions of the photon pairs $\sum^{}_{\bm{\rho}}\mathcal{C}_q(\bm{\rho})$.
%%%%%%%%%%%%%%%%%%%%%%%%%%%%%%%%%%%%%%%%%%%%%%%%%%%%
\begin{figure}[tbp]
	\centering
	\includegraphics[width=\linewidth]{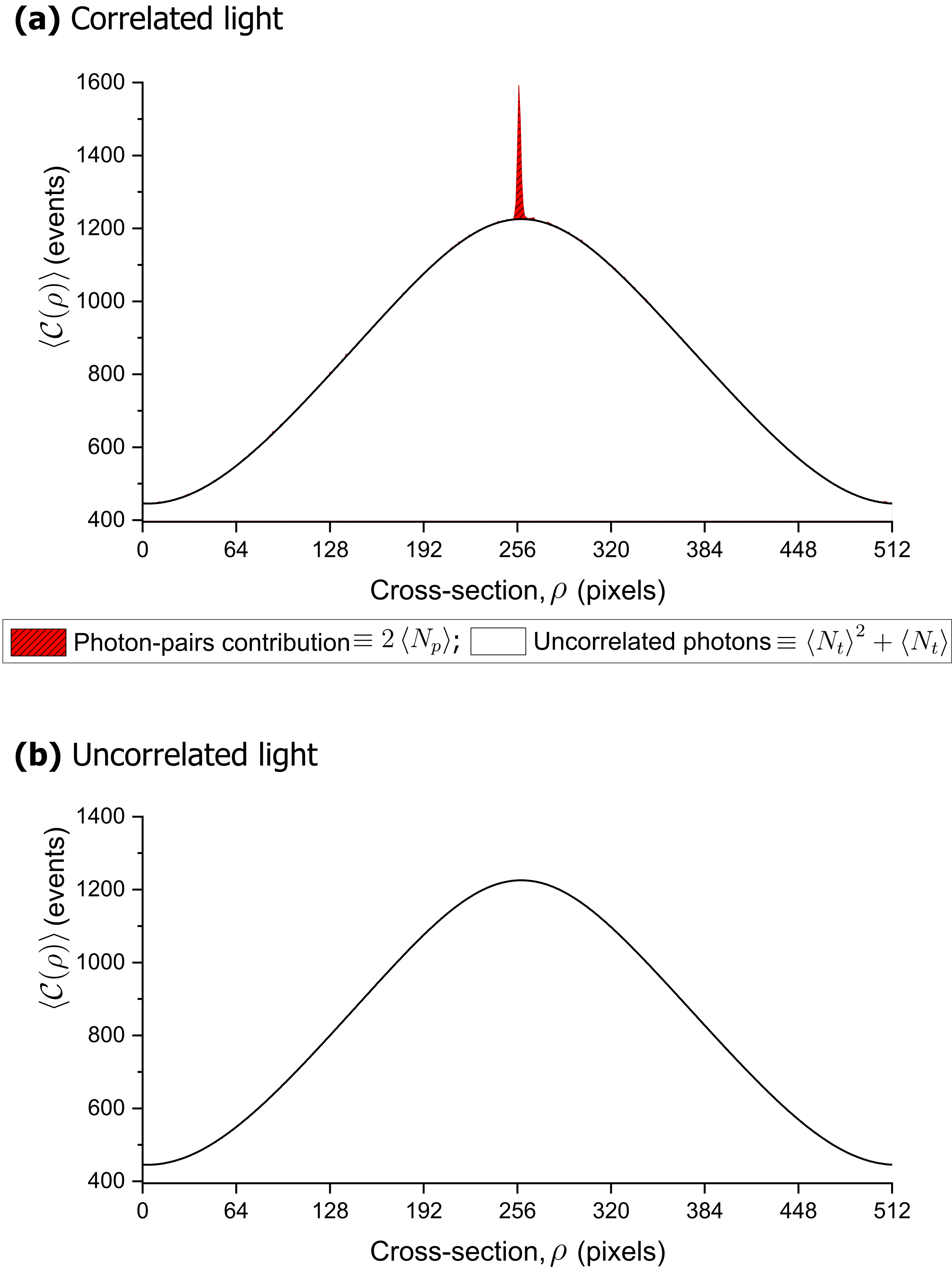}
	\caption{\textbf{Cross-sections of the autocorrelation functions for spatially correlated and uncorrelated light.} In the presence of quantum spatial correlations (a), the autocorrelation function consists of two components: a correlation peak shown in red which equals the number of jointly detected photons $2\left \langle N_p \right \rangle$, and a `pedestal' which equals the number of detected uncorrelated photons $\left \langle N_t \right \rangle^2+\left \langle N_t \right \rangle$. In the absence of quantum spatial correlations, the autocorrelation function only consists of the pedestal, as shown in (b). The cross-sections shown in (a) and (b) were fitted to experimental autocorrelation functions, which were averaged over 2500 frames acquired by our EMCCD camera.}
	\label{fig:corrUncorrRepres}
\end{figure}
%%%%%%%%%%%%%%%%%%%%%%%%%%%%%%%%%%%%%%%%%%%%%%%%%%%%
Figure~\ref{fig:corrUncorrRepres} summarises the important points demonstrated above through a schematic representation of the expected correlation function shape in both the presence and absence of correlations. The two distinguishable contributions have different spatial features: the photon pair contribution is due to the quantum correlations, it is expected to exhibit narrow spatial features since these correlation spatial extension is narrow in respect to the size of the beam, the classical contribution is due to the presence of singles, dark counts and to the fact that as there is more than one pair of photons per frame, each pair of detection events contribute in fact to the value of $\mathcal{C}$ despite not corresponding to quantum correlations. The spatial features of this last contribution is expected to be of the size of the beams and dark count shape, as any detection events contribute.\\

Finally, various figures of merit related to the correlation function can be used to choose the regime for the detection of quantum correlations. The first one is the ratio between the height of the correlation peak contribution compared to the height of the pedestal, which can be used to maximise the ability to identify actual photon pairs in images~\cite{tasca_OptimizingUseDetector_2013,edgar_ImagingHighdimensionalSpatial_2012}. The second figure of merit that can be maximised is the signal to noise ratio (SNR) of the correlation peak and the noise on the reconstructed correlation function~\cite{lantz2014optimizing}. It is desirable to use such a figure of merit when one wants to maximise the efficiency with which the pairs are detected or to characterise the correlation characteristics with a good accuracy~\cite{moreau_RealizationPurelySpatial_2012,moreau_EinsteinPodolskyRosenParadoxTwin_2014,denis_TemporalGhostImaging_2017}. Another figure of merit is the degree of correlation, which can be used to maximise the overall detection efficiency and compute a conservative value of the total effective quantum efficiency of the optical channel~\cite{heidmann_ObservationQuantumNoise_1987,brida_ExperimentalRealizationSubshotnoise_2010,toninelli_SubshotnoiseShadowSensing_2017}. In section~\ref{sec: model descr.} we discuss a simple numerical model that can be used to accurately reproduce the spatially resolved detection of quantum correlations. Moreover, the model can also be used to maximise the total effective quantum efficiency, by aiding the optimal choice of required experimental parameters.

%Finally, we propose $\left \langle N_p \right \rangle$ as a new figure of merit, which as mentioned represents the number of actually detected photon-pairs, as extracted from the cross-correlation function.

\section{Experimental setup}
\begin{figure}[tbp]
	\centering
	\includegraphics[width=\linewidth]{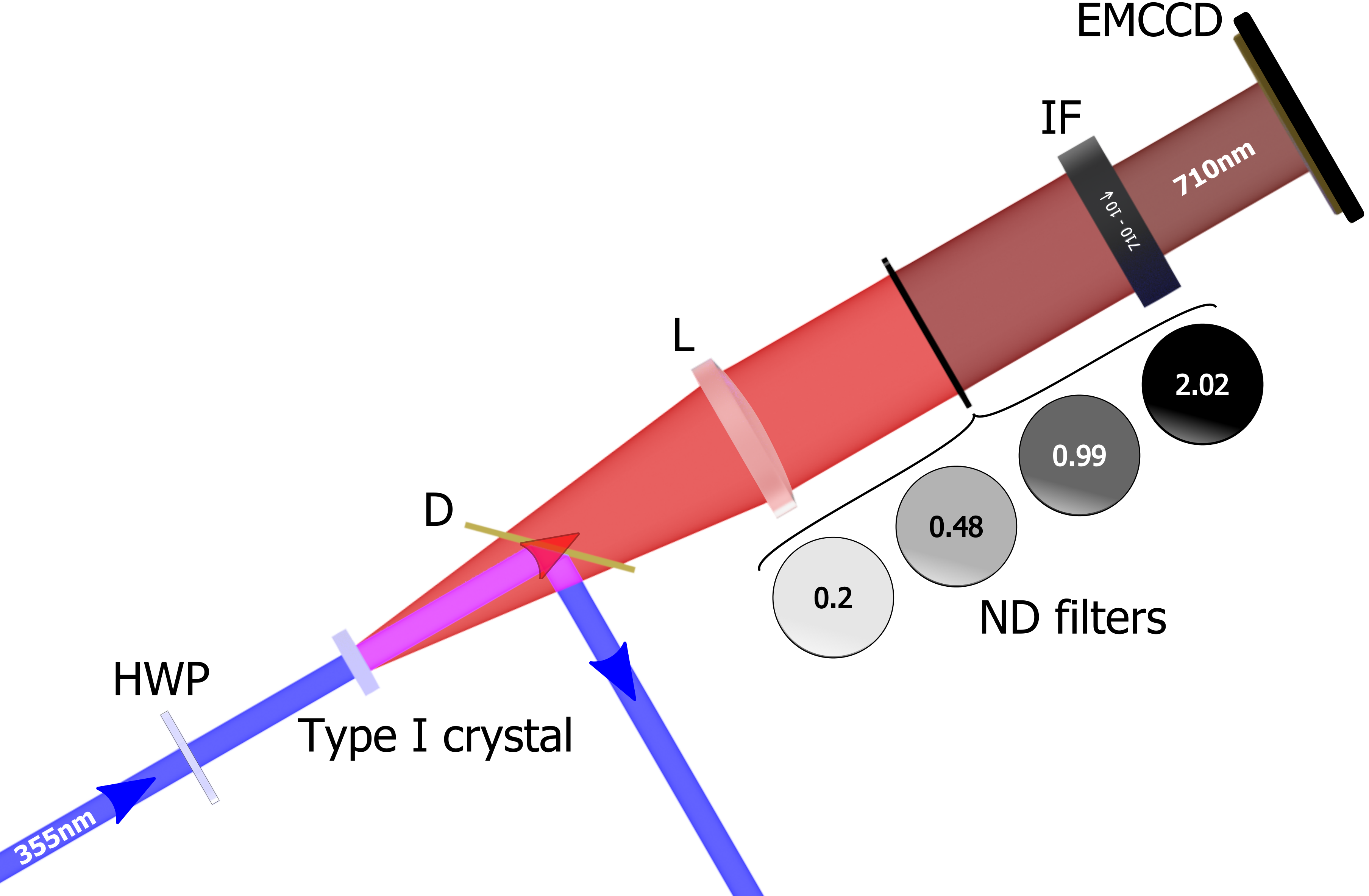}
	\caption{\textbf{Experimental apparatus.} A type-I nonlinear crystal cut for degenerate downconversion at 710nm is pumped by a 355nm UV laser, which is removed after the crystal using dichroic mirrors (D). A 250mm lens (L) is used to access the far-field of the crystal and thus the spatially anticorrelated biphotons, which are captured by the single-photon sensitive EMCCD camera and selected at degeneracy by an interference filter (IF). A combination of ND filters and a half-wave plate (HWP) is used to change the ratio of biphotons to spatially uncorrelated photons, while keeping the light flux incident on the camera at a constant level. The optical density of the employed ND filters at 710 nm, as reported by the manufacturer, is shown for convenience.}
	\label{fig:setup}
\end{figure}

The experimental set-up shown in Fig.\ref{fig:setup} was used to assess both qualitatively and quantitatively the predictions of the model, by comparing and analysing experimentally acquired binary with those produced using the model.\\
A 160 mW, 355 nm laser (JDSU, xCyte CY-355-150) is attenuated to a few mW and
used to pump a 10 mm $\times$ 10 mm $\times$ 3 mm $\beta$-barium borate (BBO) non-linear crystal, cut for type-I degenerate downconversion. Two dichroic mirrors placed after the crystal (each 98$\%$ transmissive at 710 nm) are used to remove the UV pump. An EMCCD camera (Andor, ULTRA 897, model DU897-BV) is then used to acquire images of the beam in a photon counting regime~\cite{doi:10.1046/j.1365-8711.2003.07020.x,lantz_MultiimagingBayesianEstimation_2008}. A top-hat transmission profile interference filter (IF) (centred at 710 nm, with a 10 nm square transmission band) is mounted on the camera to select the degenerate downconverted photon-pairs. We also introduce neutral density (ND) filters of different optical density after the nonlinear crystal. The average number of detected events is kept constant, by turning a zero-order half-wave plate (HWP) placed before the crystal (i.e. by changing the downconversion efficiency), thus balancing the optical loss introduced by each ND filter. The chance of jointly detecting both photons of a photon-pair is thus reduced when an ND filter is placed after the BBO crystal, causing a reduction in the number of jointly-detected photon pairs and a degradation in the strength of the detected spatial correlation. In the next section we show that the model can accurately reproduce this phenomenon and that we can estimate the value of the added optical density by extracting the ratio between spatially correlated and uncorrelated photon from the captured binary frames.
%%%%%%%%%%%%%%%%%%%%%%%%%%%%%%%%%%%%%%%%%%%%%%%%%%%%
%%%%%%%%%%%%%%%%%%%%%%%%%%%%%%%%%%%%%%%%%%%%%%%%%%%%
\subsection{Camera settings}
The EMCCD camera was operated as a photon-counting detector, by using a single discriminating threshold above which an analogue count for a pixel would be considered as one detected event. The threshold for binary detection of photons was chosen as to maximise the number of detected biphotons, using the properties of the model discussed in this paper, and as shown later in Fig.~\ref{fig:threshold}. The detailed acquisition settings are reported here: $-100^{\circ}$C cooling ($10^{\circ}$C refrigerating liquid); frame transfer enabled; 0.05775~s exposure time; 0.34483~s kinetic cycle time equal to 2.9 Hz acquisition rate; kinetics frame transfer mode; $512\times512$~pixel$^2$ frame-size; 0.5~$\mu$s vertical shift speed; 1~MHz pixel readout rate; baseline clamp on; normal clock amplitude; 1000 electron multiplying gain level; 1.0 preamplifier gain.

\section{Model description}\label{sec: model descr.}
In this section we describe the general model that can be used to generate photon-counted frames as those acquired by an EMCCD camera, illuminated by SPDC beams. For the purpose of this work in which the experimental data was produced by an EMCCD placed in the far-field of a type-I nonlinear crystal tuned for collinear and degenerate downconversion, we focus on modelling the detection of spatially anticorrelated biphotons. Therefore, in order to simulate binary frames equivalent to those acquired by the EMCCD camera shown in Fig.~\ref{fig:setup}, the model needs to produce the pixel-coordinates of detected events, which can either be bright or dark events. In this treatment we define the total number of detected events in a frame $N$ as the sum of the bright and dark events, as follows: $N=N_B+N_D$. 
%1) The overall shape of the downconverted beam switches between a Gaussian- and a sinc-like function when moving from the near-field to the far-field of the crystal~\cite{lantz2015einstein}; 2) The detected biphotons are spatially correlated and anticorrelated in the near- and far-field of the crystal respectively; 3) The autocorrelation function in the image plane of the crystal needs to be corrected by removing the central pixel, which by accounting for the correlation of all events with themselves does not contain useful information~\cite{edgar_ImagingHighdimensionalSpatial_2012}.

For what concerns dark events, the model relies on a experimentally measured noise fill-factor. This is used to set the number of randomly distributed events, thus accounting for both the thermalised photoelectrons of the EMCCD camera and its clock-induced-charge. It is also possible to use a weighted Poissonian distribution to control the fluctuations in the number of dark-events from frame to frame, thus mimicking the noise performance of the EMCCD camera, which may not operate at the shot-noise-level and is therefore affected by super-Poissonian fluctuations in the detected number of dark-events. It should be noted that for simplicity the model ignores charge-smearing effects.

For what concerns bright events, these can either be `unpaired' single-photons (which are present because of non-zero optical losses) or biphotons, which in this case are spatially anticorrelated. Therefore, the model needs to produce the pixel-coordinates of these events. Each bright-event is committed to the frame once it has been randomly checked against a user defined value for the total effective quantum efficiency, which accounts for the combined optical losses and detection efficiency of the detector.\\

More specifically, we define the distribution of detected events of coordinates $x$ and $y$ in a frame of width $d$ using a two-dimensional function $F(x,y)$ as the sum of the bright $B$ and dark $D$ events in a frame (the bottom-left corner is chosen as the origin), as follows:
\begin{align}
F(x,y)=B(x,y)+D(x,y).
\end{align}
%%%
The contribution $B(x,y)$ of the detected signal and idler photons (indicated by the $s$ and $i$ subscripts respectively) is defined as follows:
\begin{equation}\label{eqn:Bexp}
\begin{aligned}
B(x,y)=\sum_{n=1}^{N_p/\eta} \Omega(Z^n_s)\delta(x-X^n_s,y-Y^n_s)+\\
+\Omega(Z^n_i)\delta(x-X^n_i,y-Y^n_i),
\end{aligned}
\end{equation}
where $N_p/\eta$ is the number of pairs generated in the crystal, $\eta$ is a user-defined parameter for the total
effective quantum efficiency, $\delta(x,y)$ is the two dimensional delta Dirac function, $Z^n_s$ and $Z^n_i$ are random variables taking values between 0 and 1 and following a uniform distribution, and $\Omega(Z)$ is a function used to determine whether a bright event is detected or not, defined as follows:
\begin{equation}
\Omega(Z) = 
\begin{cases}
0 & \text{if $Z>\eta$ (missed event)}\\
1 & \text{if $Z\leq\eta$ (detected event)}.
\end{cases}
\end{equation}
The remaining terms $X^n_s$, $Y^n_s$, $X^n_i$, and $Y^n_i$ in Eq.~\ref{eqn:Bexp} are random variables defined by the following expressions: 
\begin{align}
X^n_s&=X_m+\frac{d}{2}\\
Y^n_s&=Y_m+\frac{d}{2}\\
X^n_i&=-X_m+X_c+\frac{d}{2}\\
Y^n_i&=-Y_m+Y_c+\frac{d}{2},
\end{align}
in which the values of the variables $X_m$, $Y_m$, $X_c$, $Y_c$ are randomly sampled from the following normal distributions:
\begin{align}
X_m&\sim\mathcal{N}(0,\sigma_m)\\
Y_m&\sim\mathcal{N}(0,\sigma_m)\\
X_c&\sim\mathcal{N}(0,\sigma_c)\\
Y_c&\sim\mathcal{N}(0,\sigma_c).
\end{align}
Moreover, $2\cdot \sigma_m$ and $2\cdot \sigma_c$ are user-defined parameters and represent the overall transverse spatial extent of the downconverted beam and the correlation-width of spatially correlated biphotons respectively. In order to simplify the mathematical description, both the modes' profile (which here represents the cross-section of the downconverted beam) and the correlation profile (which here represents the cross-section of the birth-zone in the nonlinear crystal in terms of pairs production) are approximated to Gaussian curves, as represented by the normal distributions listed above.

On the other hand, in the case of position correlated photons, equation~\ref{eqn:Bexp} still holds but the random variables defining the transverse positions of the photons need to be modified to :
\begin{align}\label{eq: position correlation}
X^n_s&=X_m+\frac{d}{2}\\
Y^n_s&=Y_m+\frac{d}{2}\\
X^n_i&=X_m+X_c+\frac{d}{2}\\
Y^n_i&=Y_m+Y_c+\frac{d}{2}.
\end{align}

Finally, the pixel coordinates of the dark-events that make up the noise contribution in a frame $D(x,y)$ are defined as follows:
\begin{equation}
D(x,y)=\sum_{n=1}^{N_D} \delta(x-X^D_n,y-Y^D_n)
\label{eqn:Dexp}
\end{equation}
where $X^D_n$ and $Y^D_n$ are random variables following a uniform distribution:
\begin{align}
X^D_n&\sim\mathcal{U}(1,d)\\
Y^D_n&\sim\mathcal{U}(1,d).
\end{align}

The model can also be readily adapted to reproduce type-II parametric downconversion, simply by separating the signal and idler photons to two different regions within the frame. As shown in Eq.~\ref{eq: position correlation}, our model can also be used to describe the detection of biphotons by a detector placed in the image plane of the crystal. In this case it is useful to note the following aspects:
\begin{enumerate}
	\item The overall shape of the downconverted beam switches between a Gaussian- and a sinc-like function when moving from the near-field to the far-field of the crystal~\cite{lantz2015einstein};
	\item The detected biphotons are spatially correlated and anticorrelated in the near- and far-field of the crystal respectively;
	\item When computing the autocorrelation function of detected frames in the image plane of the crystal, the central pixel needs to be removed as it represents the correlation of all events with themselves and takes a non-useful large intensity value~\cite{edgar_ImagingHighdimensionalSpatial_2012}.
\end{enumerate}

In practice, it is possible to account for the actual sinc-like distributions by experimentally characterising the overall intensity envelope of the downconversion beam. For example this can be done by averaging a large number of acquired bright frames, thus accounting for the specific phase-matching condition of an experiment and whether the detector is placed in the near- or far-field of the crystal. Specifically, the average of many frames can be used to extract an intensity cross-section of the detected beam, which is in turn used to define the random-number distribution for the transverse pixel-coordinates of simulated bright events.
%In its most basic implementation, our model faithfully reproduces the signal components produced by a photon-counting camera as shown in the experimental arrangement in Fig.~\ref{fig:setup}. The signal contributions are: 1) Randomly distributed events, i.e. dark-events of the detector (exposure- and readout-time dark-events and clock-induced-charge); 2) Uncorrelated bright events, distributed according to the overall shape of the downconverted beam, i.e. partially transmitted biphotons; 3) spatially correlated (in this paper anticorrelated) bright events, i.e. jointly detected photons of a biphoton-packet. The first two contributions account for the uncorrelated `pedestal` of the autocorrelation function, whereas the biphotons' contribution accounts for the $2N_p$ correlation peak, highlighted in red in Fig.~\ref{fig:corrUncorrRepres}.
Moreover, if the number of multiple detections per pixel is preserved in the computational implementation, our model can then also be used to simulate the photon-number-resolved detection of events, as later shown in Fig.~\ref{fig:deltaNp}.
\section{Model validation}
In this section we first qualitatively compare acquired frames to those produced by our model. We then quantitatively validate our model, showing that the number of detected photon-pairs $N_p$ corresponds to the number of event-pairs contained in the cross-correlation peak.\\

\begin{figure*}[tbp!]
	\centering
	\includegraphics[width=\linewidth]{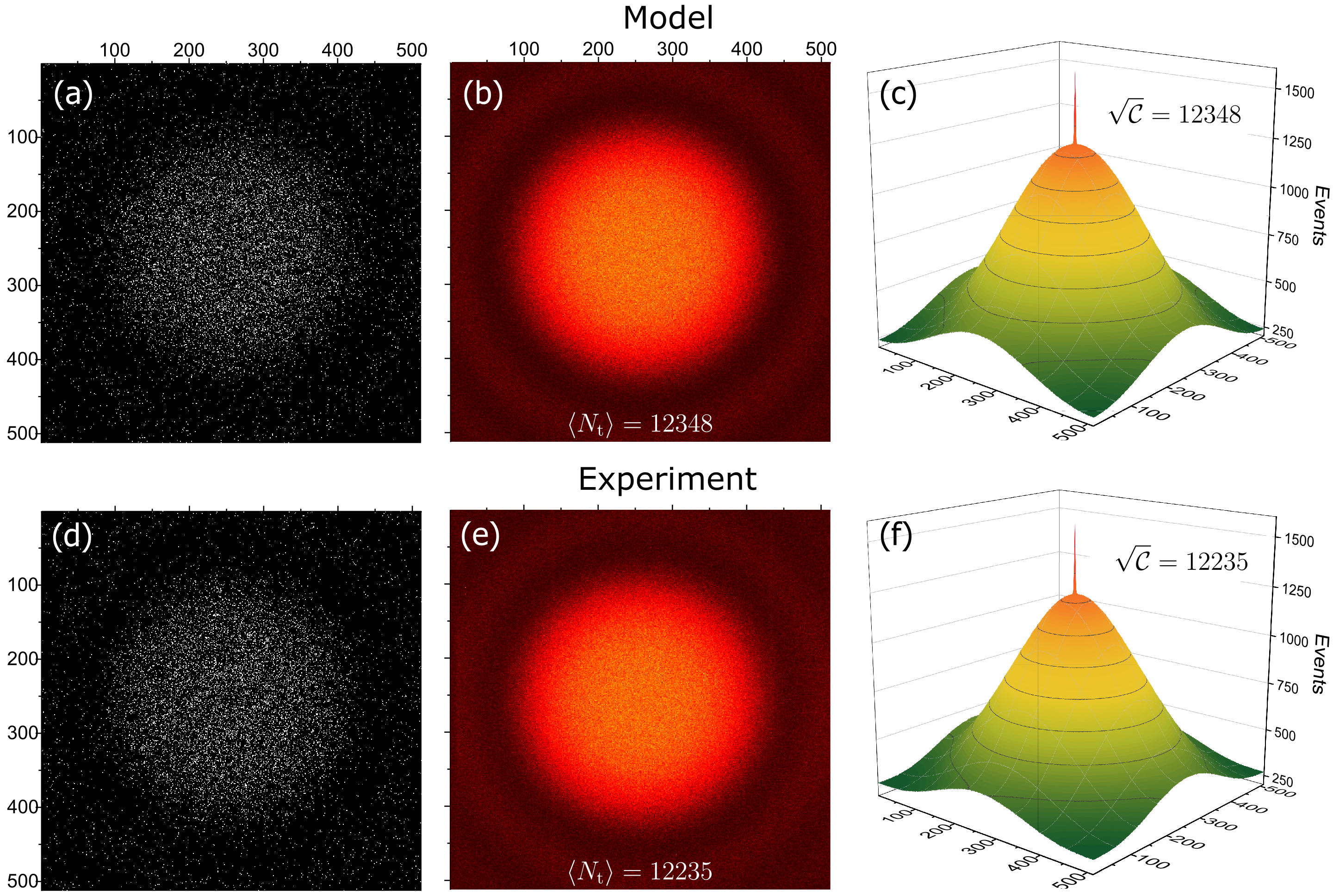}
	\caption{\textbf{Comparison of modelled and experimental detections of correlated light in the far-field of a type-I nonlinear crystal and respective cross-correlation functions.} (a) and (d) show individual binary frames, accounting for both bright and dark events; (b) and (e) shown the accumulation of 250 binary frames and report the average number of all events for one frame. (c) and (f) show the averaged cross-correlation function of 250 binary frames. The square root of the average number of correlated events $\sqrt{\mathcal{C}}$ is found to be the same as the average number of events $\left \langle N_t \right \rangle$, as predicted by Eq.~\ref{eq:fubini}.} %The overall shape of the beams can be seen to differ slightly between the model and the experiment, as visible in (b) and (b'). This is due to a simplification of the model, according to which the modes distribution in the far-field were approximated to a Gaussian distribution, instead of the sinc-like distribution visible in (b').}
	\label{fig:simExpComp}
\end{figure*}

A sample of individual frames and accumulated frames, produced by both our model and acquired using the EMCCD camera, are shown for comparison in Fig.~\ref{fig:simExpComp}(a-b) and (d-e) respectively. The average number of detected events $\left \langle  N_{\text{model}} \right \rangle$ and $\left \langle N_{\text{exp.}} \right \rangle$ for both our modelled and experimentally acquired frames are shown for convenience.
%%%%%%%%%%%%%%%%%%%%%%%%%%%%%%%%%%%%%%%%%%%%%%%%%%%%
%%%%%%%%%%%%%%%%%%%%%%%%%%%%%%%%%%%%%%%%%%%%%%%%%%%%
As expected, the total number of events in the autocorrelation function equals the average number of detected events, as predicted by Eq.~\ref{eq:fubini}. Moreover, it can be seen that the shape of the autocorrelation function for the modelled frames closely matches the experimentally computed one, shown in Fig.~\ref{fig:simExpComp}(c) and (f) respectively.

\begin{figure}[tbp!]
	\centering
	\includegraphics[width=\linewidth]{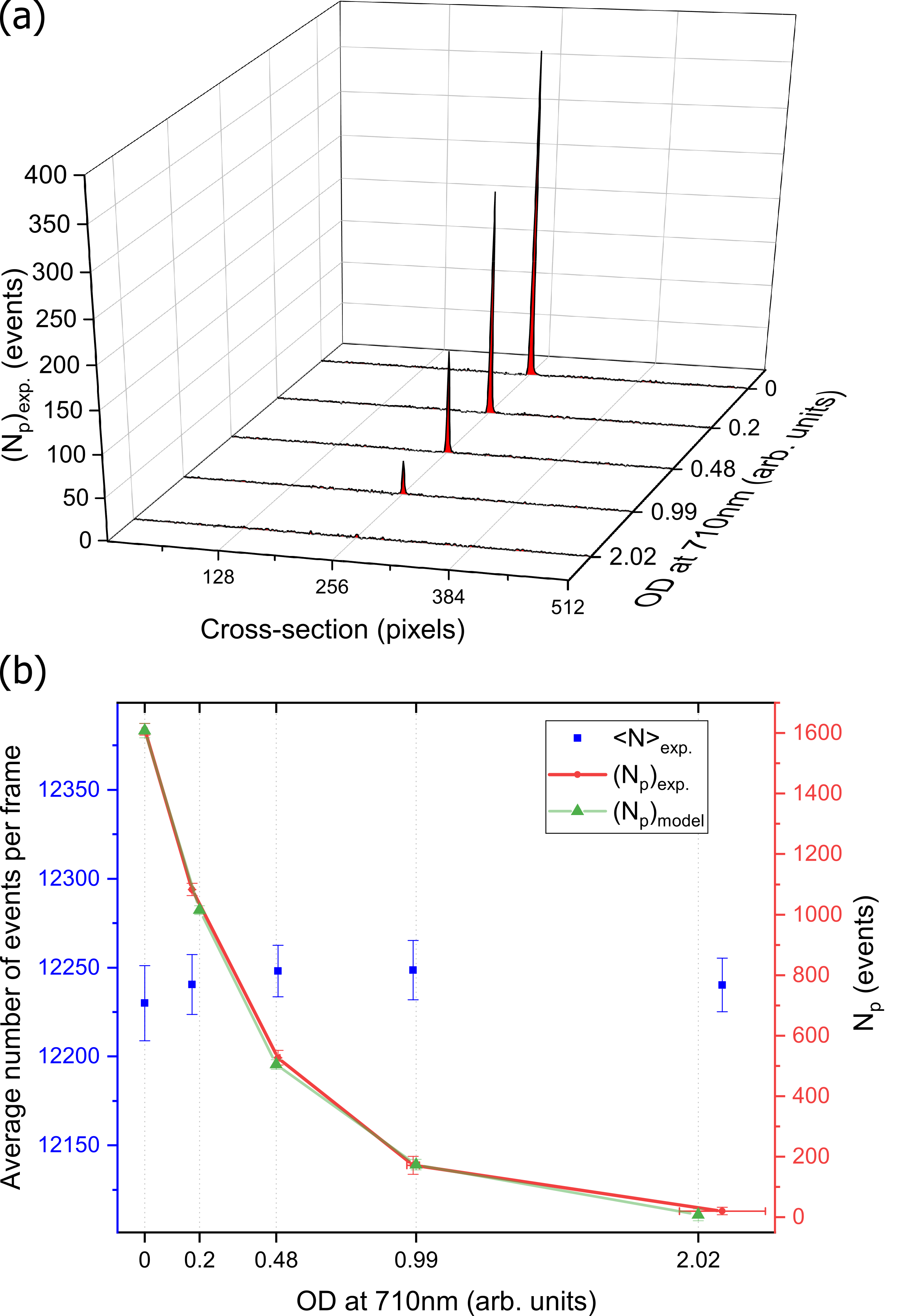}
	\caption{\textbf{Detected biphotons as a function of optical density.} The number of photon pairs $N_p$ is extracted from the averaged cross-correlation of 2500 frames per each value of optical density. (a) shows a 2D cross-section of the isolated photon-pairs form the acquired frames, obtained by subtracting the Gaussian-shaped contribution of uncorrelated events from the averaged cross-correlation of all events. By integrating the volume of these isolated correlation peaks, the number of detected photon-pairs as a function of optical density is plotted in (b), both for the acquired frames (red-series) and the modelled data (green-series). It should be noted that the number of detected photon-pairs contained in the autocorrelation peaks (red- and green-series) is found to decrease as a function of optical density, even though the average number of detected events (blue-series) is kept constant by increasing the brightness of the downconversion source.}
	\label{fig:NDsAndPeaks}
\end{figure}

For the quantitative comparison, four neutral density filters (optical densities at 710nm of 0.2, 0.48, 0.99, and 2.02) were in turn inserted after the nonlinear crystal. Importantly, the total number of detected events per frame was kept constant by increasing the downconversion efficiency, as shown in the blue-series of Fig.~\ref{fig:NDsAndPeaks}(b). Under these constant light-flux conditions we are still able to detect changes in the number of detected biphotons as caused by the added optical losses of the ND filters. As to be expected, higher levels of optical loss correspond to more photon-pairs being partially absorbed, decreasing the number of jointly-detected photon pairs in the peak of the autocorrelation function.  

A straight-forward way to extract the number detected biphotons is to subtract the uncorrelated `pedestal' from the averaged cross-correlation of all events, and then summing the remaining events. In practice, the uncorrelated pedestal can be found by curve fitting~\cite{lantz2015einstein} or background-subtraction~\cite{edgar_ImagingHighdimensionalSpatial_2012}.
Cross-sections of the averaged background-subtracted correlation peaks are shown in Fig.~\ref{fig:NDsAndPeaks}(a), whereas the integration of all correlated event-pairs contained in the peaks and for different values of optical density are shown in Fig.~\ref{fig:NDsAndPeaks}(b). The average number of detected biphotons as a function of optical density is shown in the red- and green-series of Fig.~\ref{fig:NDsAndPeaks}(b) respectively, and are reported in Table~\ref{table:NDsAndPeaks} for convenience. From these quantitative results it can be seen that the model agrees strongly with the experimental data, allowing to accurately extract the number of spatially correlated and uncorrelated events from binary frames.

%%%%%%%%%%%%%%%%%%%%%%%%%%%%%%%%%%%%%%%%%%%%%%%%%%%%
%%%%%%%%%%%%%%%%%%%%%%%%%%%%%%%%%%%%%%%%%%%%%%%%%%%%

\begin{table}[]\caption{\textbf{Estimation of optical density from the detected number of photon-pairs at constant light-level as extracted from the data plotted in Fig.~\ref{fig:NDsAndPeaks}(b).} The nominal values of optical density and transmission at 710nm for the various ND filters are reported in the first and second columns respectively. The measured number of photon-pairs for both the modelled data and the acquired frames are reported in the 3rd and 4th columns respectively. The experimentally estimated value of optical density is reported in the last column respectively. The average number of detected events and of the detector's dark events are shown at the bottom row.}
	\begin{tabular}{|c|c|c|c|c|}
		\hline
		\textbf{ND$_{710\text{nm}}$} & \textbf{T$_{710\text{nm}}$ (\%)} & \textbf{(N$_p$)$_{\text{model}}$} & \textbf{(N$_p$)$_{\text{exp.}}$} & \textbf{OD$_{\text{exp.}}$} \\ \hline
		0.00                         & 100                              & $1608\pm 24$                      & $1612\pm 19$                     & 0.00                        \\ \hline
		0.20                         & 63.3                             & $1015\pm 14$                      & $1083\pm 20$                     & $0.17\pm 0.01$              \\ \hline
		0.48                         & 33.2                             & $505\pm 16$                       & $527\pm 24$                      & $0.49\pm 0.01$              \\ \hline
		0.99                         & 25.7                             & $174\pm 17$                       & $172\pm 30$                      & $0.99\pm 0.02$              \\ \hline
		2.02                         & 10.3                             & $8\pm 19$                         & $21\pm 12$                       & $2.11\pm 0.16$              \\ \hline
	\end{tabular}
	\label{table:NDsAndPeaks}
\end{table}

Knowing the number of detected biphotons $N_p$, allows us to extract the total effective quantum efficiency $\eta$, defined in terms of the detected correlated events $\left \langle N_c \right \rangle$ and uncorrelated events $\left \langle N_u \right \rangle$ as follows:
\begin{equation}\label{eq:eta}
\eta=\frac{\left \langle N_c \right \rangle}{\left \langle N_c \right \rangle+\left \langle N_u \right \rangle}.
\end{equation}
The average number of all detected events $\left \langle N \right \rangle$ is equal to the sum of correlated $\left \langle N_c \right \rangle$ and uncorrelated $\left \langle N_u \right \rangle$ bright events, as well as the contribution due to the detector's dark events $\left \langle N_D \right \rangle$. By substituting $\left \langle N_c \right \rangle$ with the number of detected biphotons $2\left \langle N_p \right \rangle$ and using Eq.\ref{eq:eta} we have:

\begin{eqnarray}\label{eq:<N>}
\left \langle N \right \rangle&=&\left \langle N_c \right \rangle+\left \langle N_u \right \rangle+\left \langle N_D \right \rangle\\
&=&\frac{2\left \langle N_p \right \rangle}{\eta}+\left \langle N_D \right \rangle.
\end{eqnarray}

Re-arranging for $\eta$ an inserting the experimentally estimated values from Table~\ref{table:NDsAndPeaks} for zero optical density we find:
\begin{equation}\label{eq:etaCompute}
\eta=\frac{2\left \langle N_p \right \rangle}{\left \langle N \right \rangle-\left \langle N_D \right \rangle}=0.31,
\end{equation}
where $\left \langle N_p \right \rangle=1612$, $\left \langle N \right \rangle=12240$, and $\left \langle N_D \right \rangle=1985$.

\begin{figure}[tbp!]
	\centering
	\includegraphics[width=\linewidth]{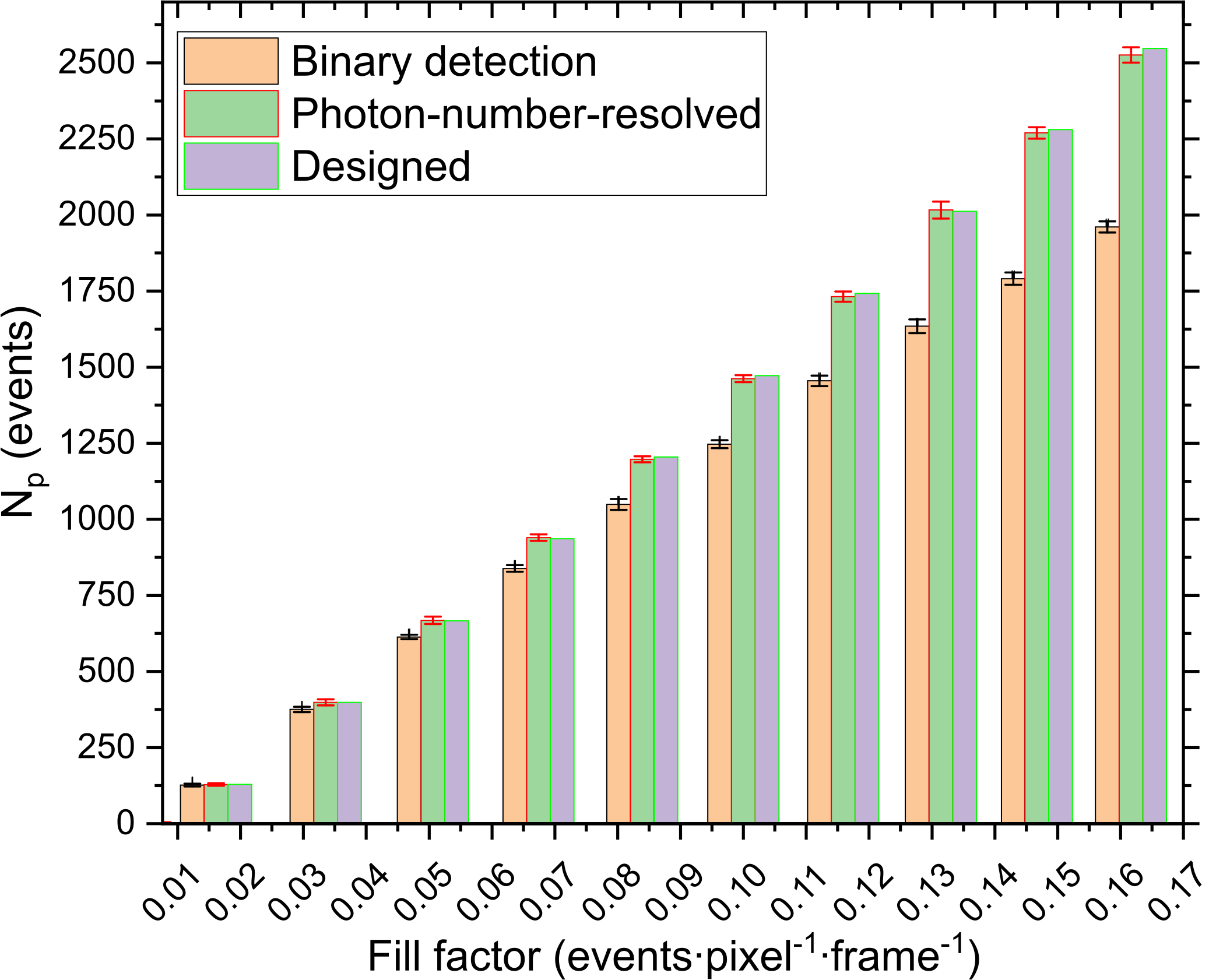}
	\caption{\textbf{Biphoton detections for a binary and an idealised photon-number-resolved camera.} Our model was used to investigate the effect of binary detection over the number of detected biphotons $N_p$, as extracted from the background-subtracted autocorrelation peak. In the case of photon-number-resolved detection (i.e. when more than one photon can be resolved per pixel) the detected number of biphotons matches the actual value used to design the synthetic frames. However, in the case of binary detection (i.e. when a pixel is allowed to detect at most 1 photon) an increasing number of biphotons is lost as the fill-factor increases. The fill-factor was computed within one standard deviation $\sigma_m\approx 142$ pixel of the downconverted beam, where roughly $68\%$ of all events are detected. }
	\label{fig:deltaNp}
\end{figure}

As it can be seen in Fig.~\ref{fig:deltaNp}, as the fill factor increases, so does the chance of multiple photons being detected by the same pixel. In the case of an idealised photon-number-resolved camera, this is not an issue, as all events are successfully accounted for. However, in the case of a binary detector (as for current state-of-the-art EMCCD cameras), high fill-factors involve loss of events, as the single discriminating threshold that is applied to the analogue photocurrents produced by each pixel means that at most only one photon per pixel can be detected. In the experimental realisation the light level was set as to not exceed $\approx 0.14$ events per pixel per frame, by computing the fill-factor within one standard deviation $\sigma_m$, corresponding to the central brightest region of the beam.

%%%%%%%%%%%%%%%%%%%%%%%%%%%%%%%%%%%%%%%%%%%%%%%%%%%%
%%%%%%%%%%%%%%%%%%%%%%%%%%%%%%%%%%%%%%%%%%%%%%%%%%%%

Having a model that perfectly mimics the spatial and quantitative properties of detected biphotons allows to extend the treatment beyond the experimentally dictated limits of binary detection, to include the case of the ideal, true photon-number resolved detection of quantum fluctuations. Thus, by enabling the detection of more than one photon per pixel it is possible to visualise the losses in detection efficiency due to the binary nature of detection. This is shown in Fig.~\ref{fig:deltaNp} in terms of detected biphotons, for both binary and photon-number resolved detections as a function of fill-factor. It can be seen that in the case of binary detection (orange-series), more and more events are lost as the fill-factor increases, due to a greater probability of events landing on the same pixel during one exposure time. By tuning the parameters of our model to the experiment and preserving multiple-event detections per pixel, we were able to produce the photon-number resolved detection of biphotons for the full range of fill-factors. The grey-series shows the actual number of biphotons that was used to generate the synthetic frames. This number matches the green-series shows the computed number of detected biphotons, found by quantifying the volume of the background-subtracted correlation peak. We therefore have confirmed the ability of or model to faithfully describe the detection of downconverted biphotons, both in the case of binary and true photon-number resolved detections.

\begin{figure}[tbp!]
	\centering
	\includegraphics[width=\linewidth]{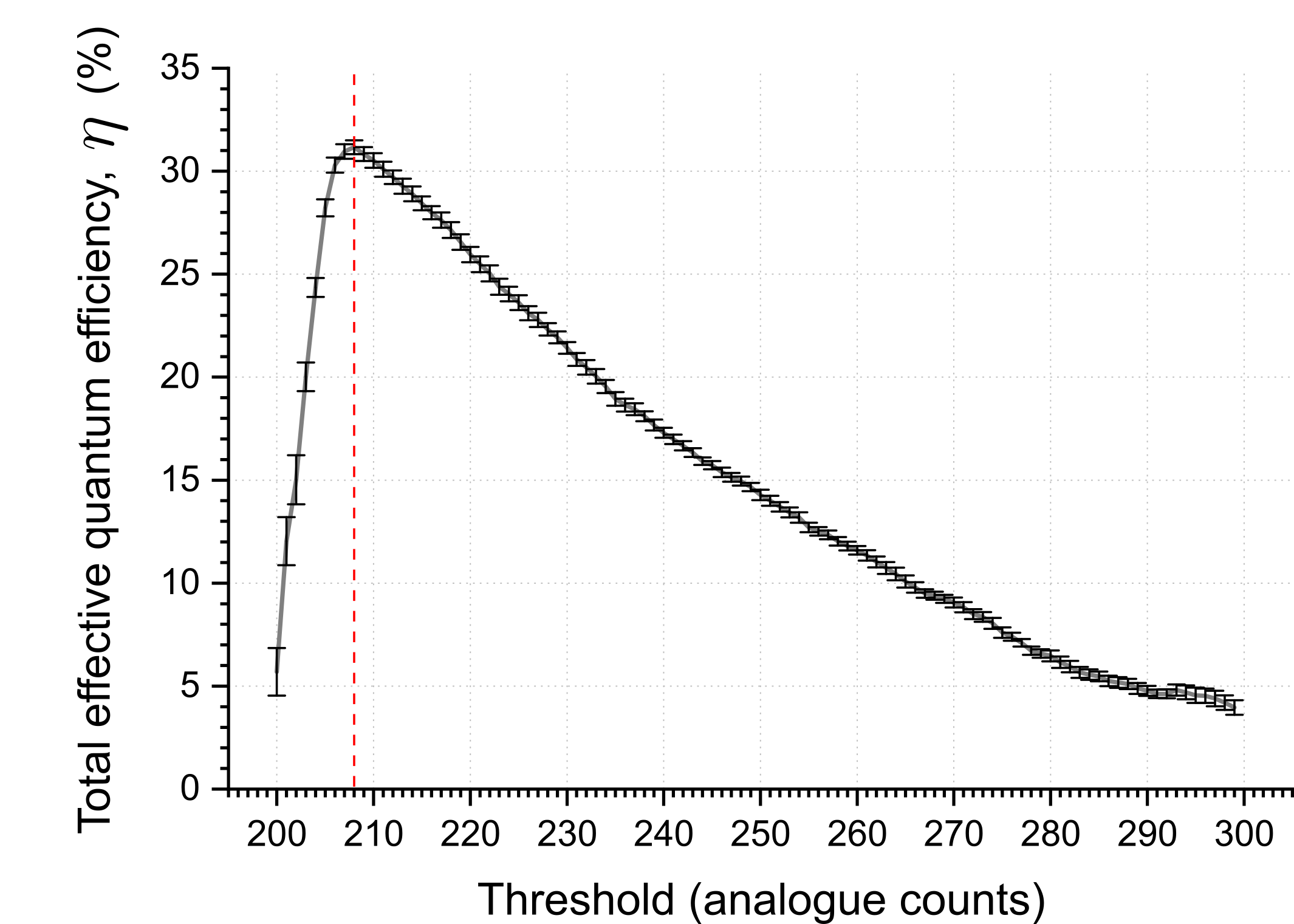}
	\caption{\textbf{Optimisation of experimental parameters aided by the estimation of $\eta$.} The properties of the autocorrelation function highlighted by the treatment of our model can be used to optimise quantum sensing and imaging experiments. In this instance, the crucial choice of the event-discriminating threshold for frames acquired with an EMCCD camera is performed using the total effective quantum efficiency $\eta$ obtained from the description of our model. The red-dotted line highlights the optimal threshold that maximises $\eta$.}
	\label{fig:threshold}
\end{figure}

As a final consideration, being able to quantify the number of detected biphotons can be a powerful diagnostic tool for the optimisation of quantum imaging and sensing schemes, where small effects such as defocus or tilting of interference filters can negatively affect the overall quantum enhancement. Hence, it is possible to iteratively fine-tune each component, using the average number of detected photon-pairs as a metric for performance. To illustrate this concept, we show how even the choice of the threshold in analogue counts for an EMCCD camera can be guided using the total number of detected biphotons and thus the derived total effective quantum efficiency of the detection scheme, as shown in Fig.~\ref{fig:threshold}.

%%%%%%%%%%%%%%%%%%%%%%%%%%%%%%%%%%%%%%%%%%%%%%%%%%%%
%%%%%%%%%%%%%%%%%%%%%%%%%%%%%%%%%%%%%%%%%%%%%%%%%%%%

\section{Conclusion}
In this work we have presented a simple numerical model that can reproduce both the binary and photon-number-resolved detections of biphotons produced by a nonlinear crystal. Importantly, we have drawn an analogy between the mathematical expression of the intensity correlation function for spatially detected quantum fluctuations and the spatial cross-correlation function of a frame with a 180-degree rotated copy of itself. From this description we have shown how the contribution due to quantum correlations (i.e. spatially correlated events) can be extracted from the autocorrelation function, and used as a metric for the strength of detected correlations and related total effective quantum efficiency. By comparing the analysis of modelled frames to those acquired using an EMCCD camera we have confirmed the model's ability to accurately reproduce both quantitative and spatial properties of downconverted photons, as detected by a photon-counting camera. Our model can therefore be used to accurately simulate quantum sensing and imaging experiments based on the detection of spatially correlated photons, as well as aiding the optimisation of experimental parameters, including detector settings, which can have an effect on the detection efficiency and thus the attainable quantum-enhancement.

\textbf{Acknowledgements}
E.T. acknowledges the financial support from the EPSRC Centre for Doctoral Training in Intelligent Sensing and Measurement (EP/L016753/1). P.-A. M. acknowledges the support from the European Union's Horizon 2020 research and innovation programme under the Marie Sklodowska-Curie fellowship grant agreement No 706410, of the Leverhulme Trust through the Research Project Grant ECF-2018-634 and of the Lord Kelvin / Adam Smith Leadership Fellowship scheme. T.G. acknowledges the financial support from the EPSRC (EP/N509668/1) and the Professor Jim Gatheral quantum technology studentship. This work was also funded by the UK EPSRC (QuantIC EP/M01326X/1).

%\bibliography{correlationsPaper}

\end{document}